\documentclass[12pt, a4paper]{article}

\usepackage[margin=1in]{geometry}
\usepackage{graphicx}
\usepackage{amsmath}
\usepackage{amssymb}
\usepackage{booktabs}
\usepackage{tabularx}
\usepackage{longtable}
\usepackage{array}
\usepackage{hyperref}
\usepackage[authoryear, round]{natbib}
\usepackage{url}
\usepackage{float}
\usepackage{enumitem}
\usepackage{authblk}

\hypersetup{
    colorlinks=true,
    linkcolor=blue,
    citecolor=blue,
    urlcolor=blue
}

\title{From Horizontal Layering to Vertical Integration: A Comparative Study of the AI-Driven Software Development Paradigm}

\author[1,$\dagger$]{Chi Zhang}
\author[1,2,$\dagger$,*]{Zehan Li}
\author[1]{Ziqian Zhong}
\author[1,2,3,*]{Haibing Ma}
\author[1]{Dan Xiao}
\author[1]{Chen Lin}
\author[4]{Ming Dong}

\affil[1]{Moximize.ai, Shanghai, China}
\affil[2]{AI Committee, China Creative Studies Institute}
\affil[3]{Shanghai Zhongqiao Vocational and Technical University, Shanghai, China}
\affil[4]{Shanghai Jiaotong University, Shanghai, China}
\affil[$\dagger$]{These authors contributed equally to this work}
\affil[*]{Corresponding author e-mail: LZH@moximize.ai, mhb@shzq.edu.cn}

\date{}

\begin{document}

\maketitle

\begin{abstract}
This paper examines the organizational implications of Generative AI adoption in software engineering through a multiple-case comparative study. We contrast two development environments: a traditional enterprise (brownfield) and an AI-native startup (greenfield). Our analysis reveals that transitioning from Horizontal Layering (functional specialization) to Vertical Integration (end-to-end ownership) yields 8$\times$ to 33$\times$ reductions in resource consumption. We attribute these gains to the emergence of Super Employees, AI-augmented engineers who span traditional role boundaries, and the elimination of inter-functional coordination overhead. Theoretically, we propose Human-AI Collaboration Efficacy as the primary optimization target for engineering organizations, supplanting individual productivity metrics. Our Total Factor Productivity analysis identifies an AI Distortion Effect that diminishes returns to labor scale while amplifying technological leverage. We conclude with managerial strategies for organizational redesign, including the reactivation of idle cognitive bandwidth in senior engineers and the suppression of blind scale expansion.
\end{abstract}


\section{Introduction}

\subsection{Research Background: From Efficiency Gains to Paradigm Shift}

The integration of Generative AI (GenAI) into software engineering extends beyond tool upgrades to fundamentally alter the production function of software development. We observe this transformation unfolding across four progressively deepening layers. At the most immediate level (Layer 1), AI agents demonstrate unprecedented capabilities in code generation, with studies quantifying efficiency gains of 30 to 100 times in specific tasks; however, such task-level efficiency does not automatically translate to organizational velocity if surrounding structures remain unchanged. This necessitates a shift in development paradigms (Layer 2), where the dramatic reduction in coding costs renders traditional Horizontal Layering (Functional Silos) obsolete. Consequently, the industry is shifting towards AI-Driven Vertical Integration, where AI-augmented Super-Cells manage end-to-end delivery, reinterpreting Conway's Law by radically reducing the number of necessary communication nodes. Yet, this velocity introduces deep-seated risks (Layer 3); specifically, when AI generates code faster than humans can verify, a Human-in-the-Loop mechanism becomes the non-negotiable safety valve for liability and architectural integrity. Ultimately, these shifts culminate in a societal and organizational transformation (Layer 4), characterized by the rise of the Super Employee and the Atomic Organization. This paradigm allows companies to scale output without proportionally scaling headcount, fundamentally altering the Total Factor Productivity (TFP) equation by suppressing the reliance on labor scale.

\subsection{Research Problem}

While Layer 1 (Efficiency) is well-documented, a critical gap remains in understanding the organizational restructuring (Layer 2) required to harness it. Current structures, rooted in \citet{taylor1911} scientific management and operationalized through assembly-line logic, constrain the productivity potential of AI. These legacy paradigms prioritize process standardization over cognitive integration, creating organizational friction that limits GenAI's effectiveness.

This study investigates how software organizations must restructure, from Horizontal Layering to Vertical Integration, to maximize Human-AI Collaboration Efficacy.

\subsection{Research Contributions: Three Levels of Innovation}

This paper presents a multiple-case comparative study, analyzing both a traditional enterprise (Case A) and an agile AI-native team (Case B), with contributions structured across three distinct levels. First, regarding paradigm innovation, we define the structural shift from Horizontal Layering (Functional Teams) to Vertical Integration (Cross-Functional Super-Cells), visualizing the collapse of silos into end-to-end ownership. Second, we propose a standard innovation wherein the Maximization of Human-AI Collaboration Efficacy serves as a new primary optimization metric for engineering management. This metric supplants simplistic outputs, such as lines of code, with a focus on how effectively human judgment directs AI execution. Finally, we provide an actionable strategy and managerial playbook for this transition. This includes reactivating idle cognitive bandwidth in senior engineers, suppressing blind scale expansion in favor of talent density, and shifting the primary driver of TFP from labor scale and management efficiency to AI-driven technology.

\section{Literature Review}

We organize literature into four thematic streams: the evolution of software development paradigms, the empirical impact of GenAI on coding efficiency and quality, the theoretical foundations of organizational structure, and the dynamics of human-AI collaboration.

\subsection{The Evolution of Software Development Paradigms}

The organization of software development has historically mirrored the industrial principles of the era.

\citet{smith1937} established the economic foundation for the Division of Labor. \citet{taylor1911} subsequently operationalized these principles into a management science, advocating for task decomposition, workflow standardization, and the separation of planning from execution. Ford's assembly line scaled this logic industrially in 1913, establishing the organizational paradigm that persists in software engineering today. In software engineering, this industrial logic manifested as the Horizontal Layering model. Mirroring Taylor's decomposition of tasks, software production was sliced into distinct functional silos: frontend, backend, testing, and operations. This structure aims to optimize local efficiency but inadvertently introduces high communication costs, acting as the handover friction described in Conway's Law.

\citet{conway1968} provides the theoretical link between these organizational structures and the software systems they produce, positing that systems design is constrained by the communication structures of the organization. Historically, methodologies like Waterfall and later Agile and DevOps attempted to optimize these communication paths. While Agile improved feedback loops and DevOps bridged the gap between development and operations, they largely retained the fundamental division of labor, focusing on optimizing the handover rather than eliminating the need for it.

Recent research suggests that GenAI challenges this Smithian foundation. By acting as a versatile tool that spans multiple domains, AI enables a shift towards less fragmented, more integrated workflows, potentially rendering traditional silos obsolete \citep{hou2024}. This shift aligns with the emerging concept of Agentic AI \citep{wang2025}. \citet{durante2024}, in their survey ``Agent AI: Surveying the Horizons of Multimodal Interaction,'' describe a transition from passive tools to interactive agents capable of perception, planning, and execution across multimodal domains. This agentic capability allows for the creation of Super-Cells, where a single human-agent unit can handle complex, cross-functional tasks that previously required a team, effectively collapsing the horizontal layers.

\subsection{GenAI in Software Engineering: The Efficiency Frontier}

The immediate impact of GenAI on software engineering is a dramatic increase in coding efficiency, often referred to as the Layer 1 impact.

\citet{yang2024} empirically demonstrated that advanced LLMs (like GPT-4 and GLM-4) can achieve efficiency improvements ranging from 30 to 100 times over traditional manual coding in specific contexts. This empirical evidence supports the broader economic analysis by OpenAI researchers. \citet{eloundou2024}, in their seminal paper \textit{``GPTs are GPTs,''} categorize Large Language Models (LLMs) as General-Purpose Technologies (GPTs). They estimate that LLMs could affect at least 10\% of tasks for 80\% of the US workforce, with software engineering being among the most highly exposed sectors. This implies that the efficiency gains observed by \citet{yang2024} are not isolated anomalies but indicators of a systemic labor market shift.

Further supporting this, \citet{hou2024} surveyed the landscape of Large Language Models (LLMs) for software engineering, identifying that LLMs are not merely code completion tools (like early Copilot versions) but are evolving into autonomous agents capable of requirements engineering, code generation, and testing.

However, this efficiency is not uniform. \citet{becker2025} found in a randomized controlled trial that while AI tools are expected to boost productivity, they can sometimes slow down experienced developers in complex, open-source environments if the tasks require deep contextual understanding that the AI lacks. This nuances the efficiency narrative, suggesting that the gain is highly dependent on the nature of the task and the organizational context.

\subsection{Code Quality and Reliability in AI-Assisted Development}

As barriers to code generation decrease, research attention shifts to the quality and reliability of AI-generated artifacts.

\citet{ray2025} formally discusses the emerging phenomenon of Vibe Coding, a concept originating from developer communities (often associated with the intuitive, trial-and-error interaction with LLMs). Ray describes this as a new mode of development where practitioners rely on the general intent of the output rather than rigorous syntactic verification. While this accelerates prototyping, \citet{ray2025} warns of the long-term maintainability risks and the potential for a QA crisis if developers lose the ability to debug the code they effectively commissioned but did not write.

\citet{naqvi2025} provides a rigorous qualitative evaluation of LLM-generated versus human-written code. Their findings indicate that while LLMs can produce code that is syntactically correct and often functionally adequate, they introduce subtle, high-risk vulnerabilities (hallucinations) that are distinct from human errors. This underscores the necessity of a robust verification layer, aligning with our proposed Human-in-the-Loop (HITL) mechanism.

The practical application of such validation mechanisms is explored by \citet{lizh2025} and \citet{ma2025} who introduced the TEMAI framework, demonstrating how multimodal AI can be integrated into real-world workflows with rigorous evaluation standards. Their findings suggest that while AI can handle execution, the definition of success standards and final validation remains a critical human function, reinforcing the HITL necessity in high-stakes environments.

\subsection{Organizational Implications: From Scale to Density}

The economic implications of AI in software production are analyzed through the lens of \textbf{Total Factor Productivity (TFP)}.

\citet{vanbeveren2012} and recent analyses by \citet{acemoglu2025} suggest that AI's impact on TFP is driven by its ability to substitute for labor in routine tasks while complementing high-skill decision-making. This supports our hypothesis of the AI Distortion Effect, where the contribution of labor scale (L) is suppressed in favor of technological capability (T).

\citet{mckinsey2025} and \citet{sakthivel2025} describe the emergence of the Agentic Organization, where workflows are reimagined around AI agents. They argue for a shift from hierarchical structures to flat networks of empowered teams, a concept that mirrors our Super-Cell model. \citet{kwan2011} previously revisited Conway's Law to show that task-level alignment is crucial; in the AI era, this implies that if the task is handled by an AI-augmented individual, the organization must shrink to match that single node to avoid unnecessary complexity.

Recent economic analysis provides a theoretical mechanism for this structural flattening. \citet{brynjolfsson2025} utilize empirical data to demonstrate an upward compression of skills, where AI disproportionately boosts the productivity of lower-skilled workers, effectively narrowing the performance gap between novices and experts. Expanding on this, \citet{lij2024} introduces the Great Compression framework, arguing that AI enables a sideway compression where a single AI-augmented individual can outperform larger, traditional teams. Li posits that by allowing individuals to span both thinking (Yogi) and doing (Commissar) roles, AI reduces the need for specialized coordination managers, thereby validating the shift towards the leaner, high-density Super-Cells proposed in this study.

\subsection{Cognitive Dynamics: The Shift to Supervision}

Finally, literature addresses the cognitive shift required of human engineers.

Cognitive Load Theory \citep{sweller1988} posits that human working memory is limited. Traditional development often saturated this memory with low-level syntax (Intrinsic Load). \citet{lee2025} found that while GenAI reduces the effort of information gathering, it increases the cognitive demand for critical verification and task stewardship.

This aligns with the findings of \citet{haque2025}, who note that developers experience different types of cognitive load: less on syntax recall, more on architectural integration. This supports our Cognitive Bandwidth Optimization theory, where the Super Employee operates at a high-level supervisory capacity, minimizing Idle Bandwidth waste.

\section{Theoretical Framework: The New Development Paradigm}

We deconstruct the paradigm shift through three core insights: structural transformation, role redefinition, and the shift in value creation. Furthermore, we introduce the primary optimization metric of Human-AI collaboration and analyze organizational complexity using Conway's Law and the Raptor Engine metaphor.

\subsection{Core Insights of the AI-Driven Paradigm}

The integration of Generative AI necessitates a fundamental reconstruction of how software development organizations are designed and how value is created.

\subsubsection{Structural Transformation: From Horizontal Layering to Vertical Integration}

Traditionally, software engineering organizations have adopted a \textbf{Horizontal Layering} (Functional) structure. This model divides labor based on technical specializations, Frontend, Backend, Data, Algorithms, Testing, and Operations: to manage the complexity of large-scale systems. While this approach promotes specialization, it introduces significant communication overhead and handover latency between layers.

\begin{figure}[H]
    \centering
    \includegraphics[width=0.9\textwidth]{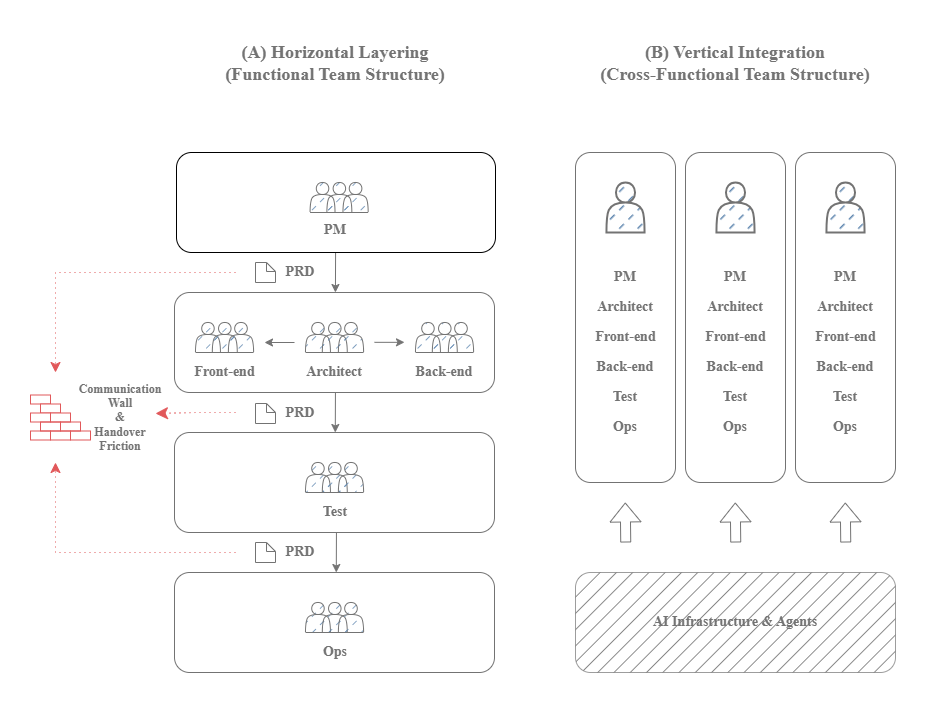}
    \caption{Comparison of Organizational Structures. Left: Traditional Horizontal Layering (Functional Teams) with high handover costs. Right: AI-Driven Vertical Integration (Cross-Functional Super-Cells) enabling end-to-end ownership.}
    \label{fig:org_structure}
\end{figure}

In the AI-driven paradigm, the organizational structure shifts towards \textbf{Vertical Integration} (Cross-Functional). With AI agents handling the implementation details across the full stack, the technical barriers between layers dissolve. A single engineer, augmented by AI, can now manage the entire link from requirement analysis to deployment. This shift moves the organization from a Functional Silo model to a Super-Cell model, where individual nodes possess complete functional capabilities (Figure \ref{fig:org_structure}).

\subsubsection{Role Redefinition: From Coder to Architect and Supervisor}

The role of human engineer undergoes qualitative change. In the traditional paradigm, a significant portion of an engineer's time is dedicated to the execution of code, translating logic into syntax. In the new paradigm, this execution role is offloaded to AI.

The human role consequently evolves into a composite of three functions:

\begin{itemize}
    \item \textbf{Architect:} Designing the high-level system structure and defining the logic boundaries.
    \item \textbf{Supervisor:} Managing AI agents, reviewing their output, and correcting hallucinations or logical errors.
    \item \textbf{Liability Holder:} Bearing the ultimate responsibility for the system's reliability, security, and correctness. This necessitates a Human-in-the-Loop (HITL) mechanism where human judgment serves as the final gatekeeper.
\end{itemize}

\subsubsection{Value Creation Shift: Maximizing Cognitive Bandwidth}

The core mechanism of value creation shifts from Individual Efficacy in execution to \textbf{Human-AI Collaboration Efficacy} in decision-making.

We propose that AI adoption enables Cognitive Bandwidth Optimization, the reallocation of engineers' mental resources from routine execution to high-value decision-making. In the traditional paradigm, senior engineers accumulate cognitive capacity that remains underutilized as routine coding tasks consume progressively less of their capability. This underutilization constitutes Idle Bandwidth Waste.

\begin{figure}[H]
    \centering
    \includegraphics[width=0.9\textwidth]{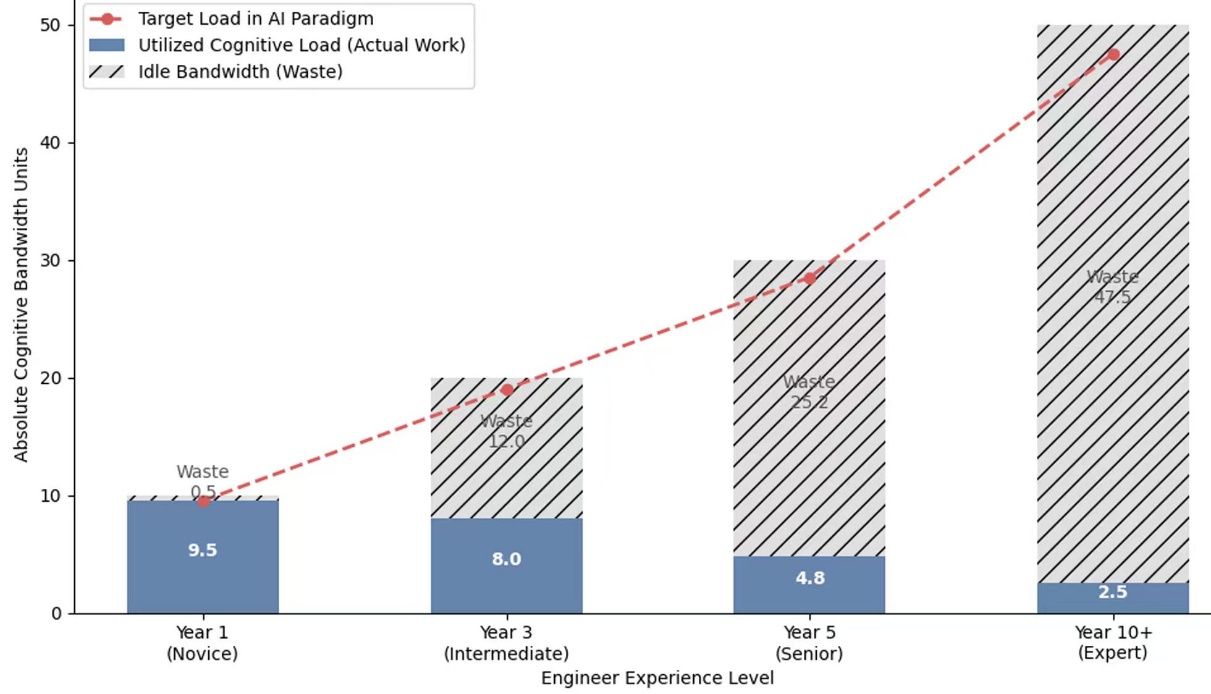}
    \caption{Cognitive Bandwidth Evolution Chart. A bar chart illustrating the data from Table \ref{tab:cognitive_evolution}. The chart serves as a visual heuristic to demonstrate the widening gap (Idle Bandwidth) between Total Capacity and Utilized Load in the traditional paradigm.}
    \label{fig:cognitive}
\end{figure}

The AI paradigm reactivates this idle bandwidth. By automating execution tasks, AI enables senior engineers to operate at sustained high cognitive engagement, focusing on architecture, judgment, and decision-making, thereby maximizing capacity utilization throughout their career lifecycle (Table \ref{tab:cognitive_evolution} and Figure \ref{fig:cognitive}).

\begin{table}[H]
\centering
\caption{Evolution of Cognitive Load and Resource Waste in Traditional Paradigm (Conceptual Model)}
\label{tab:cognitive_evolution}
\begin{tabular}{p{2.5cm}p{2.5cm}p{2.5cm}p{2cm}p{4cm}}
\toprule
\textbf{Tenure Stage} & \textbf{Total Cognitive Capacity (Units)} & \textbf{Actual Utilized Load (Units)} & \textbf{Idle Bandwidth (Waste)} & \textbf{Remarks} \\
\midrule
Year 1 (Novice) & 10 (Baseline) & 9.5 (95\% Utilization) & 0.5 & High load, low output; learning curve steep. \\
Year 3 (Intermediate) & 20 & 8.0 (40\% Utilization) & 12 & Entering comfort zone; routine tasks increase. \\
Year 5 (Senior) & 30 & 5.0 ($\sim$16\% Utilization) & 25 & High experience, but significant time spent on repetitive execution. \\
Year 10+ (Expert) & 50 (5x Growth) & 2.5 (5\% Utilization) & 47.5 & Massive Resource Waste; high capability underutilized on implementation details. \\
\bottomrule
\end{tabular}
\begin{flushleft}
\small\textit{Note: Data presented in Table \ref{tab:cognitive_evolution} derives from a theoretical conceptual model developed through expert consensus during the study's qualitative interviews. Values are illustrative estimations intended to visualize the trend of relative cognitive resource allocation, rather than precise psychometric measurements.}
\end{flushleft}
\end{table}

\subsection{Human-AI Collaboration and Organizational Complexity}

To evaluate the effectiveness of this new paradigm, we establish a new normative standard and apply complexity theory to organizational design.

\subsubsection{The Primary Optimization Metric: Maximization of Human-AI Collaboration Efficacy}

The traditional functional division of labor is becoming obsolete. We propose that the primary optimization metric for the new paradigm is the \textbf{Maximization of Human-AI Collaboration Efficacy}.

Unlike static job descriptions, the collaboration boundary between humans and AI is flexible, dynamic, and personalized to the specific organization. The ultimate criterion for this boundary is not whether a task \textit{can} be automated, but whether the specific assignment of tasks between humans and AI maximizes the total system output per unit of human cognitive investment.

\subsubsection{Measuring Organizational Complexity: Conway's Index}

We utilize Conway's Law to interpret the structural constraints, while drawing on Brooks' Law to quantify the communication overhead. While \citet{conway1968} links design to communication structure, \citet{brooks1975} defines the intercommunication complexity (C) using the formula:

\begin{equation}
C = \frac{N(N - 1)}{2}
\end{equation}

Where N represents the number of distinct nodes (teams or specialized departments) required to deliver a feature.

In the traditional Horizontal Layering model, N is high (Frontend + Backend + Test + Ops + Product), leading to quadratic complexity (C). The AI-driven Vertical Integration model drastically reduces N (often to 1 or small integers), thereby reducing C significantly.

However, while organizational complexity decreases, the responsibility and cognitive complexity borne by the single human node increase. This shift aligns with the Cognitive Bandwidth theory, pushing the human node to operate at the frontier of their capability boundaries.

\subsubsection{The Raptor Engine Metaphor: Simplification via Integration}

\begin{figure}[H]
    \centering
    \includegraphics[width=0.85\textwidth]{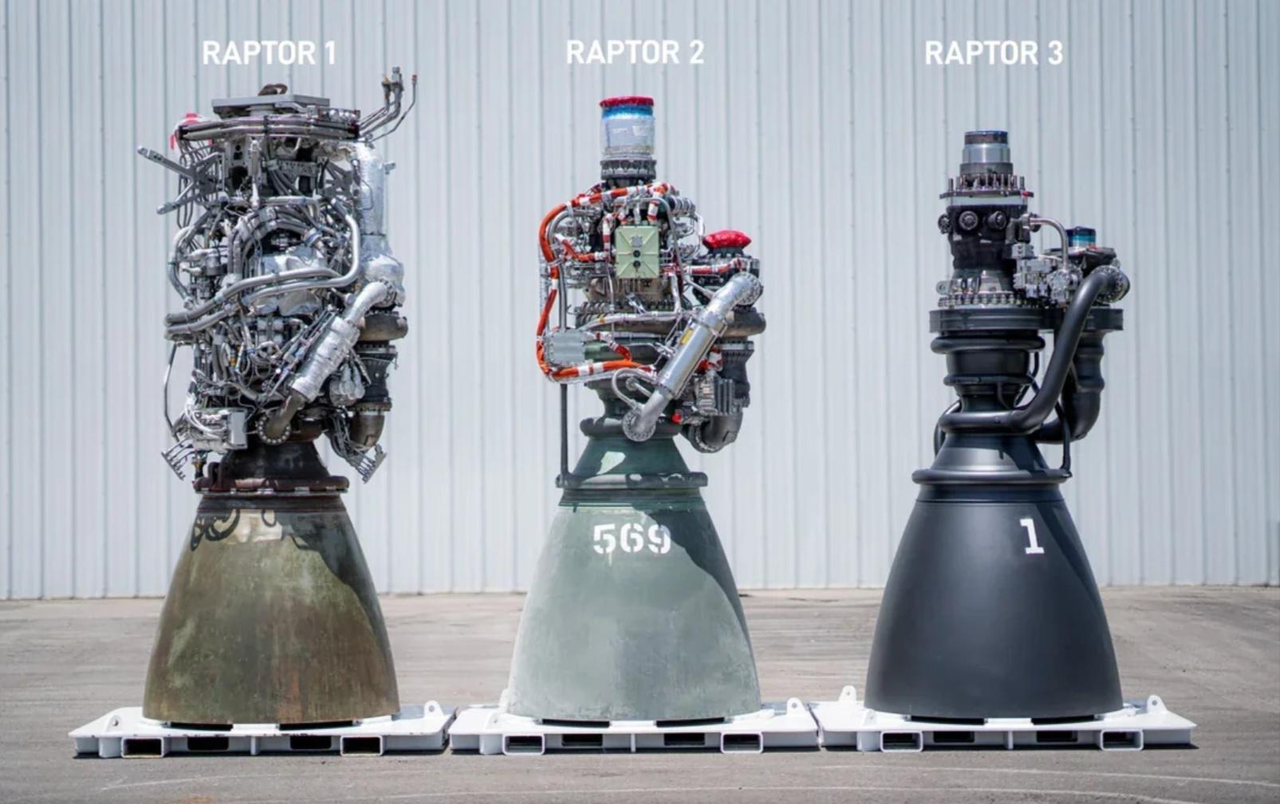}
    \caption{The structural evolution of SpaceX Raptor engines (Raptor 1 to Raptor 3) illustrating radical simplification. The merging of complex subsystems into an integrated whole serves as a heuristic for the transition from siloed software teams to AI-augmented end-to-end ownership. Source: SpaceX (2024).}
    \label{fig:raptor}
\end{figure}

The evolution of the SpaceX Raptor Engine provides a useful analogy (Figure \ref{fig:raptor}). The progression from Raptor 1 (complex tubing, redundant components) to Raptor 3 (integrated, minimalist design) mirrors the organizational simplification enabled by AI. Just as Raptor engineers improved thrust-to-weight ratio by merging subsystems and eliminating intermediate piping, the AI software paradigm eliminates organizational intermediaries, including PRDs, handovers, and coordination layers, to achieve more direct value delivery (Table \ref{tab:raptor}). This analogy draws on the public documentation of SpaceX's engineering evolution \citep{spacex2024}.

\begin{table}[H]
\centering
\caption{Comparative Analogy: Raptor Engine Evolution vs. AI Software Paradigm}
\label{tab:raptor}
\small
\begin{tabular}{p{2.2cm}p{3.5cm}p{3.5cm}p{4cm}}
\toprule
\textbf{Dimension} & \textbf{Traditional Paradigm (The Complex Engine)} & \textbf{AI-Driven Paradigm (The Raptor Model)} & \textbf{Impact} \\
\midrule
Structural Complexity & High Redundancy: Multiple pipes, disparate components, complex routing. & Radical Simplification: Merged channels, reduced parts, direct flow. & Eliminates structural friction and maintenance overhead. \\
Organizational Equivalent & Siloed Roles: Frontend / Backend / Testing / Ops. Heavy handovers. & End-to-End Ownership: Merged roles, simplified process, removal of intermediaries & Reduces Intercommunication Overhead (C). \\
Process Interface & Document-Driven: PRD/MRD as necessary but lossy connectors. & Direct Execution: Intent-to-Code. Intermediate documentation interfaces disappear. & Direct transmission of intent; reduced information loss. \\
Result & Fragility: High management cost; single-point failures in communication. & Anti-Fragility: Efficiency leap; easier maintenance; higher stability. & Efficiency Leap: Order-of-magnitude improvement in delivery speed. \\
\bottomrule
\end{tabular}
\end{table}

\section{Research Methodology}

Given the nascent and transformative nature of Generative AI in software engineering, this study adopts a multiple-case comparative research design. Qualitative inquiry is particularly appropriate when the boundaries between the phenomenon and its context are not clearly evident, and the research seeks to answer ``how'' and ``why'' questions regarding contemporary events over which the researcher has little or no control \citep{yin2018}.

This study aims to explore the mechanisms of the AI-Driven Development Paradigm in real-world settings, favoring deep empirical evidence over broad statistical generalization. A comparative case study allows for a deep dive into the complex interplay between organizational structure, human agency, and AI technology across different contexts, offering robust evidence for theory building \citep{eisenhardt1989}.

\subsection{Research Design: The Triadic Structure across Dual Cases}

To isolate the effects of the new paradigm, we established a three-part research structure involving a Research Partner, an Execution Subject, and a Verification Scenario. This structure allows us to observe the transmission and execution of the paradigm in controlled yet authentic commercial environments across two distinct cases.

\begin{itemize}
    \item The Research Partner (Organization B): Served as the transformation catalyst, providing the theoretical framework, the Vertical Integration organizational model, and the AI-native toolchain guidance.
\end{itemize}

We applied this structure to two contrasting Execution Subjects to verify the paradigm's universality:

\subsubsection{Case A: The Traditional Enterprise (Company G - Project W)}

\begin{itemize}
    \item \textbf{Execution Subject (Company G's Seed Team):} Company G is an established traditional software company possessing over a decade of history in customized software development and rigid management processes. A Seed Team was selected from their existing workforce to act as the pilot unit for the paradigm shift. This selection controls for variable talent quality, allowing us to attribute performance gains to the paradigm shift rather than individual outliers.
    \item \textbf{Verification Scenario (Project W):} A complex business management and data acquisition system chosen as the validation field. Unlike a trivial toy app, Project W requires high reliability and integration, making it a robust testbed for validating the feasibility of End-to-End Ownership in a production environment.
    \item \textbf{Significance:} Represents the Brownfield scenario, transforming legacy structures and mindsets in a mature organization.
\end{itemize}

\subsubsection{Case B: The AI-Native Startup (Organization B - Internal AI-CRM)}

\begin{itemize}
    \item \textbf{Execution Subject (Organization B's AI Team):} An AI-native team born in the GenAI era, operating with extreme agility and no legacy debt.
    \item \textbf{Verification Scenario (Internal AI-CRM Project):} An agentic customer relationship management system designed from the ground up using AI-first principles.
    \item \textbf{Significance:} Represents the Greenfield scenario, demonstrating the theoretical limit of efficiency when legacy constraints are minimized.
\end{itemize}

\subsection{Comparative Study Approach}

This research employs a comparative study approach, contrasting the \textit{actual} outcomes of the AI-driven paradigm against the \textit{projected} outcomes of the traditional paradigm.

Since commercial projects are real-world undertakings, we cannot run a simultaneous control group developing the exact same software. Instead, we utilize counterfactual analysis based on historical baseline data and standard industry estimation models (e.g., function point analysis).

We compare two distinct states for each case:

\begin{itemize}
    \item \textbf{The Traditional Baseline (Counterfactual):} The projected resource allocation, organizational layering (frontend/backend/test), and timeline required to deliver the project using standard Waterfall/Agile methodologies.
    \item \textbf{The AI-Driven Reality (Actual):} The actual resource consumption, organizational structure (Vertical Integration), and timeline achieved using the AI-driven paradigm.
\end{itemize}

\subsection{Data Collection: Triangulation}

To ensure the validity and reliability of our findings, we employed data triangulation, collecting evidence from three distinct sources to corroborate facts and mitigate observer bias \citep{patton2014, yin2018}.

(1) Immersive Participant Observation

We acted as embedded observers within the collaboration process, capturing real-time dynamics, including shifting human-AI collaboration boundaries and immediate cognitive load experiences, that retrospective accounts might miss.

(2) Semi-Structured Interviews

We conducted in-depth interviews with key stakeholders, including the Seed Team members (executors), Company G's management (strategic sponsors), and AI engineers from Organization B. These interviews focused on the subjective experience of role transition (from Coder to Architect) and the perception of responsibility.

(3) Project Documentation and Artifacts

We analyzed objective artifacts, including:

\begin{itemize}
    \item \textbf{Git Commit Logs and Codebases:} To verify the proportion of AI-generated code and the architectural integrity.
    \item \textbf{Resource Management Reports:} To quantify person-month data for efficiency comparison.
    \item \textbf{Meeting Minutes and Communication Logs:} To trace the reduction in communication nodes (N) and the elimination of intermediate documentation (e.g., PRDs).
\end{itemize}

\subsection{Dimensions of Analysis}

Aligning with our Theoretical Framework, we analyze the data across three primary dimensions:

\begin{itemize}
    \item \textbf{Structural Transformation:} Comparing the Horizontal Layering of the traditional baseline against the Vertical Integration of the execution teams. We measure this through the reduction in functional departments and handover nodes.
    \item \textbf{Role Redefinition:} Analyzing how the daily activities of the team members shifted. We look for evidence of the transition from execution-heavy coding to high-cognitive architecture and supervision tasks.
    \item \textbf{Value Creation Shift (Human-AI Collaboration Efficacy):} Quantifying the efficiency gains (Input vs. Output). We specifically look for order-of-magnitude differences in delivery speed and cost, validating the theoretical shift from Individual Efficacy to Collaboration Efficacy.
\end{itemize}

\section{Case Analysis: Cross-Case Evidence}

To validate the universality of the AI-Driven Development Paradigm, we present empirical evidence from two contrasting cases: Project W (Case A) and Internal AI-CRM (Case B). By comparing a traditional enterprise with an AI-native team, we demonstrate that the efficiency gains and structural shifts are consistent across different organizational contexts.

\subsection{Case A: Project W (The Traditional Enterprise)}

\subsubsection{Project Background}

Project W was initiated by Company G, a traditional software company, as a foundational component of its digital transformation strategy. The system is defined as a Business Management and Data Acquisition Platform, designed to eliminate data silos and integrate disparate business processes across the group. Unlike a simple prototype, Project W is a complex enterprise-grade system requiring high concurrency handling, strict data security, and seamless integration with legacy ERP systems.

The project execution followed the \textbf{Triadic Structure} defined in our methodology:

\begin{itemize}
    \item Research Partner (Organization B): Provided methodological guidance on the Vertical Integration organizational model and the AI-native toolchain.
    \item Execution Subject (The Seed Team): A Seed Team selected from Company G's existing workforce. These engineers were traditional developers (frontend/backend specialists) who underwent immersive training to transition into AI-Augmented Architects.
    \item Validation Scenario (Project W): The shared battlefield where the new paradigm was tested against rigid business requirements.
\end{itemize}

As a strategic executive from Company G noted in an interview:

\begin{quote}
``We recognized that our traditional competitive advantage based on accumulated experience was diminishing. We needed to embrace AI not merely as a tool, but as a strategic imperative.''
\end{quote}

\subsubsection{The Traditional Baseline (Counterfactual)}

To establish a baseline for comparison, Company G conducted a rigorous resource estimation for Project W using their standard industry metrics (Function Point Analysis) and historical data from similar past projects. Under the Traditional Paradigm (Horizontal Layering), the project plan required:

\begin{itemize}
    \item \textbf{Specialized Roles:} A standard composition of Product Managers, UI/UX Designers, Frontend Developers, Backend Developers, Testers, and DevOps Engineers.
    \item \textbf{Process Efficiency:} A waterfall or agile process involving multiple handovers.
    \item \textbf{Estimated Effort:} The total effort was calculated at approximately 100 Person-Months (e.g., a team of $\sim$15 specialists working for $\sim$6.5 months).
\end{itemize}

This high estimate stems from the structural complexity inherent in horizontal layering, what we term the Communication Tax (Conway's Complexity). As noted in our interviews:

\begin{quote}
``In the traditional mode, PRDs (Product Requirement Documents) are lossy compression of information. A Product Manager spends days writing a document that captures only 10\% of the actual intent, and the developer spends days decoding it, often incorrectly.'' --- Research Partner
\end{quote}

\subsubsection{The AI-Driven Reality (Actual)}

Company G's Seed Team executed Project W using the AI-Driven Paradigm (Vertical Integration).

\begin{itemize}
    \item \textbf{Actual Effort:} Approximately 12 Person-Months (Executed by 4 engineers over 3 months).
    \item \textbf{Efficiency Gain:} $\sim$8.3x reduction in resource consumption.
    \item \textbf{Observation:} The massive specialized team was replaced by a compact Super-Cell unit, where each individual was responsible for the end-to-end link (UI to Logic to DB).
\end{itemize}

Reflecting on the organizational shift, the executive added:

\begin{quote}
``The traditional organization division is based on Taylor's scientific management\ldots treating people as tools. Now, the human function has elevated; they have become architects, managers, supervisors, and judges.''
\end{quote}

\subsection{Case B: Internal AI-CRM (The AI-Native Team)}

\subsubsection{Project Background}

This internal project by Organization B's AI Team involved building a next-generation CRM system based on an Agentic Framework. Unlike traditional CRMs, this system utilizes AI agents to autonomously handle leads processing and workflow orchestration.

\subsubsection{The Traditional Baseline (Counterfactual)}

Standard development for a system of this complexity (Customer Management, Sales Pipeline, Agentic Orchestration) would typically require a full cross-functional team.

\begin{itemize}
    \item \textbf{Standard Team Composition:} 1 Architect + 1 Product Manager + 1 Frontend + 2 Backend + 1 QA/Ops = 6 people.
    \item Estimated Duration: $\sim$8 months.
    \item \textbf{Estimated Effort:} $\sim$50 Person-Months (including communication overhead).
\end{itemize}

\subsubsection{The AI-Driven Reality (Actual)}

The project was executed by using the new paradigm.

\begin{itemize}
    \item \textbf{Actual Effort:} $\sim$1.5 Person-Months (1 Engineer, 1.5 Months).
    \item \textbf{Efficiency Gain:} $\sim$33x reduction in resource consumption.
    \item \textbf{Observation:} These observed efficiency gains suggest a potential upper bound for the new paradigm when organizational friction is near-zero (Greenfield context). As the Lead Engineer of Case B reflected:
\end{itemize}

\begin{quote}
``In this new paradigm, I don't feel like I'm writing software; I feel like I'm directing a team of experts. My role shifted from `how to build' to `what to build' and `is it correct'. The speed is exhilarating, but the cognitive load of maintaining context across the full stack is intense.''
\end{quote}

\subsection{Cross-Case Synthesis: The New Paradigm}

The comparison of Case A and Case B reveals a consistent pattern of structural collapse and efficiency explosion. Table \ref{tab:cross_case} summarizes these findings.

\begin{table}[H]
\centering
\caption{Comparative Analysis of Traditional vs. AI-Driven Paradigms (Cross-Case)}
\label{tab:cross_case}
\small
\begin{tabular}{p{1.8cm}p{3.5cm}p{3.5cm}p{4.5cm}}
\toprule
\textbf{Dimension} & \textbf{Traditional Paradigm (Horizontal Layering)} & \textbf{AI-Driven Paradigm (Vertical Integration)} & \textbf{Comparison of Impact} \\
\midrule
Structure & Functional Silos: Frontend / Backend / Test / Ops separated. & Super-Cell: One engineer manages the full link (UI to Logic to DB). & Case A: Silos merged into seed units. Case B: Native super-cell structure. \\
Process & Waterfall/Agile: Multiple handovers; PRD as the interface. & End-to-End: Intent-to-Code directly; PRD disappears as an interface. & Both cases eliminated intermediate documentation, relying on direct Agent interaction. \\
Role & Specialized Executor: Focus on syntax translation. & Architect and Supervisor: Focus on judgment and liability. & Shift from Writing to Reviewing is consistent across both cases. \\
Efficiency & High Resource Cost: Case A: $\sim$100 PM; Case B: $\sim$50 PM & Low Resource Cost: Case A: $\sim$12 PM (8x); Case B: $\sim$1.5 PM (33x) & Efficiency gains range from 8x to 33x, validating the Order-of-Magnitude hypothesis. \\
Key Issues & Information loss during handovers; slow iteration. & Dynamic Instability: Rapid tool obsolescence; high cognitive load. & The primary challenge shifts from Coordination to Cognitive Management. \\
\bottomrule
\end{tabular}
\end{table}

\subsection{Risk Exclusion: Keep Humans in the Loop}

A critical finding from case analysis is that while AI executes the work, it cannot assume liability. The dramatic increase in efficiency introduces a new risk: The Black Box Effect, where code is generated faster than it can be comprehended.

To mitigate this, the project mandated a strict \textbf{Human-in-the-Loop} protocol.

\begin{itemize}
    \item \textbf{Role Upgrade:} The engineer's primary duty shifted from writing code to reviewing architecture and audit.
    \item \textbf{Liability Assignment:} Every line of AI-generated code required human sign-off. The human architect serves as the discriminator against AI hallucinations.
\end{itemize}

As vividly described by a Seed Team member during the retrospective interview:

\begin{quote}
``Previously, I spent approximately nine hours writing code and thirty minutes on planning and design. Now the ratio has inverted: I spend most of my time on architecture and logic reviews. The cognitive demands are intense, but the value generated per hour has increased substantially. AI produces the majority of the code; I retain full ownership and responsibility for the outcome.''

--- Seed Team Engineer (Semi-structured Interview)
\end{quote}

This validates that the new paradigm does not make humans lazy; rather, it forces the utilization of all of their cognitive bandwidth for high-stakes decision-making, ensuring that the Raptor Engine of AI productivity remains under strict human control.

\section{Discussion: Re-thinking Division of Labor and the Future of Organizations}

The empirical findings from Project W and Internal AI-CRM extend beyond mere efficiency gains; they signal a fundamental restructuring of the economic logic governing software production.

\subsection{Revisiting Industrial Paradigms}

The traditional software development paradigm has long relied on the principles of \citet{smith1937} Division of Labor and \citet{taylor1911} Scientific Management. Later operationalized by Henry Ford's assembly line, this logic dictates that efficiency maximizes when complex processes are broken down into standardized, discrete tasks performed by specialized workers.

However, the AI-driven paradigm suggests a shift away from the assembly-line model. When AI assumes the role of the universal executor (handling standard coding, testing, and deployment), the human need not specialize in a single fragment of the process. Instead, the human role elevates to a Generalist-Ruler, managing the entire lifecycle. This marks a return to the Master Craftsman model, but one amplified by industrial-scale AI capabilities. This represents a transition from specialized, sequential workflows to integrated, AI-augmented delivery models.

\subsection{The Emergence of the Super Employee}

This paradigm shift gives rise to a new talent archetype: the Super Employee \citep{zhang2025}. Based on our case observations and broader market trends, we classify this archetype into two distinct categories:

\subsubsection{Type I: The Multi-Role Generalist (Width Expansion)}

Type I represents the expansion of capability boundaries. A single individual performs functions previously requiring multiple distinct roles.

\begin{itemize}
    \item \textbf{Analogy:} The Key Opinion Leader (KOL) creator economy. Supported by mature infrastructure (e.g., TikTok, YouTube), one creator handles scripting, filming, editing, and operations, tasks that formerly required a film crew.
    \item \textbf{Software Context:} An engineer who autonomously handles product design, coding, testing, and deployment. This type is prevalent among freelancers and early-stage startups where agility outweighs scale.
\end{itemize}

\subsubsection{Type II: The Exponential Efficiency Specialist (Depth Amplification)}

Type II represents the exponential amplification of output within a specific domain.

\begin{itemize}
    \item \textbf{Definition:} An engineer who retains their primary function (e.g., core systems development) but utilizes AI to increase their code output and problem-solving velocity by 10x or more.
    \item \textbf{Corporate Reality:} Our observation of Company G suggests that Type II is the dominant model for enterprise transformation. Large organizations prioritize the order-of-magnitude efficiency gains over purely blurring role boundaries. However, Type II employees inevitably experience some Type I expansion as they take ownership of adjacent tasks (e.g., a backend engineer writing frontend connector via AI).
\end{itemize}

The ultimate goal is the fusion of Type I and Type II, individuals who possess both the broad architectural vision to span domains and the AI-amplified velocity to execute them.

\subsection{Dynamics of Total Factor Productivity (TFP)}

To rigorously quantify the impact of AI, we analyze the \textbf{Total Factor Productivity (TFP)} function, drawing on the foundational framework established by \citet{solow1957}, and reinterpret it through the lens of recent AI macroeconomics \citep{acemoglu2025}. Building on the theoretical foundation established in Section 3, we now apply TFP analysis to interpret our empirical findings. TFP represents the efficiency with which inputs are converted into outputs, typically expressed as:

\begin{equation}
TFP = \frac{Y}{K^{\alpha} \times L^{(1 - \alpha)}}
\end{equation}

Where:

\begin{itemize}
    \item $Y$ = Total Output
    \item $K$ = Capital Input
    \item $L$ = Labor Input (Scale)
    \item $\alpha$ = Capital's share of output (typically 0.3--0.4)
\end{itemize}

The empirical findings from Project W and Internal AI-CRM extend beyond mere efficiency gains; they signal a fundamental restructuring of the economic logic governing software production. Traditionally, TFP growth is driven by three sources: Efficiency Improvements (Management), Scale Effects, and Technological Progress. Our research suggests that AI does not contribute to these factors uniformly. Instead, it fundamentally alters their weights:

(1) Technology as the Key Driver

AI acts as a pulse of technological progress, directly boosting output (Y) without a proportional increase in Labor (L).

(2) Management Synergy

AI positively reinforces management efficiency by reducing the complexity of coordination (lowering Conway's N). However, we maintain the Management weight at 30\% because the reduction in coordination quantity is offset by the increased demand for decision quality, shifting the managerial focus from logistical scheduling to high-stakes architectural supervision and liability governance.

(3) Suppression of Scale Effects

Crucially, AI suppresses the contribution of the Scale Factor (L). In the pre-AI era, scaling output required scaling headcount (Scale Effect). In the AI era, adding more humans to a Super-Cell structure may yield diminishing returns or even negative utility due to re-introduced coordination costs.

In the traditional paradigm (A), scaling output requires scaling headcount. In the AI paradigm (B), adding more humans to a Super-Cell structure may yield diminishing returns due to re-introduced coordination costs. Organizations are thus incentivized to remain Atomic.

\begin{figure}[H]
    \centering
    \includegraphics[width=0.9\textwidth]{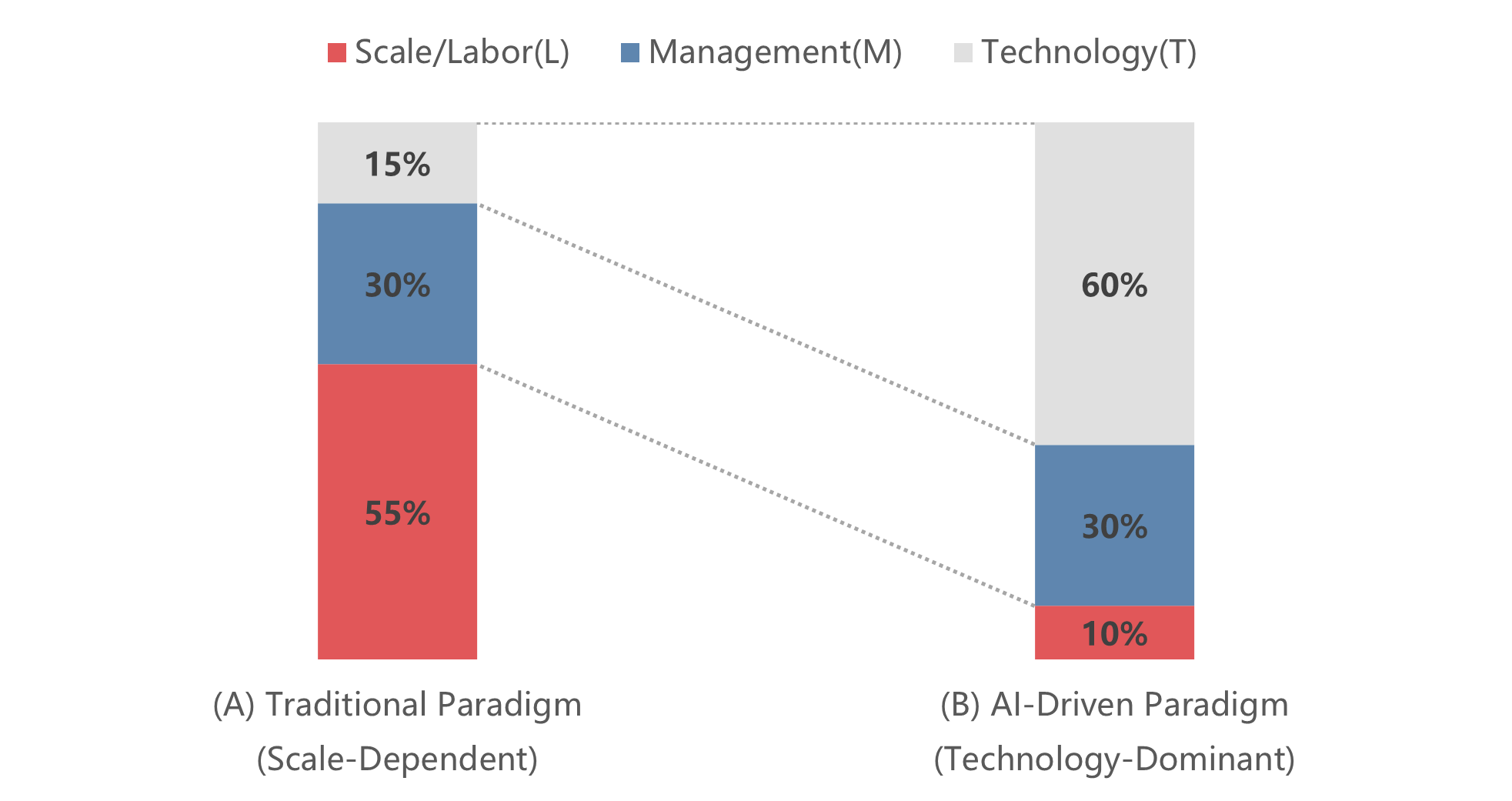}
    \caption{The Changing Weights of TFP Components in the AI Era. (A) Traditional Paradigm: Heavily reliant on Scale/Labor. (B) AI Era: Dominated by Technology, with Scale contribution significantly diminished. Values are developed through expert consensus during the study's qualitative interviews.}
    \label{fig:tfp}
\end{figure}

\subsection{The Flexible Organization}

A critical implication of the Super Employee model is the move towards Flexible Organizations.

Unlike the standardized bricks of the Ford assembly line, where every worker was replaceable and roles were fixed, the AI-driven organization is organic. Because each individual's ability to leverage AI (their Collaboration Efficacy) varies, and their domain expertise differs, the optimal organizational shape becomes unique to the specific composition of its talent.

(1) From Static Hierarchy to Dynamic Configuration

Our findings suggest that organizations may benefit from transitioning away from rigid org charts toward dynamic pods or Super-Cells that form around specific deliverables. Rather than maintaining permanent functional departments, resources can be assembled into full-stack units defined by capability density rather than headcount. A single Type II Super Employee may deliver what traditionally required an entire department.

(2) Reconsidering Standardized Job Descriptions

In the Taylorist model, the role defines the person; in the AI paradigm, the person increasingly defines the role. Given that output variance between average engineers and AI-augmented Super Employees can exceed 10$\times$, standardized job levels may lose utility. Our findings suggest value in personalized role design, where responsibility scope matches individual cognitive bandwidth and AI fluency rather than conforming to predefined classifications.

(3) Outcome-Oriented Governance

As internal processes within Super-Cells become less transparent due to AI-mediated execution, traditional process metrics (hours worked, lines of code, ticket volume) lose diagnostic value. Our analysis indicates that organizations may achieve better results through outcome-oriented governance, defining clear deliverable standards at the interface level while granting Super-Cells autonomy over execution methods.

\subsection{Risks and Dynamic Evolution}

While the potential is immense, the transition is fraught with challenges that organizations must navigate:

\begin{itemize}
    \item \textbf{Ethical and Moral Risks:} The Black Box nature of AI code generation raises concerns about hidden biases or vulnerabilities. If an AI system causes harm, the moral and legal burden falls heavily on the Human Supervisor, creating high psychological pressure.
    \item \textbf{Workforce Displacement:} The suppression of the Scale Effect implies reduced demand for junior execution-focused roles, creating potential structural unemployment for developers unable to transition to Type I or Type II Super Employee configurations.
    \item \textbf{Dynamic Instability:} It is crucial to acknowledge that this methodology is dynamically evolving. As AI models update (e.g., from GPT-5 to future iterations), the optimal boundary between human and machine will shift. The primary optimization metric is not a static line but a moving target that requires continuous calibration.
\end{itemize}

\section{Conclusion}

This study investigated the structural and economic implications of Generative AI integration in software engineering through a multiple-case comparative analysis. Our findings provide empirical evidence that AI adoption necessitates not merely tool upgrades but comprehensive organizational paradigm shifts.

\subsection{Principal Findings}

The central contribution of this research is the empirical validation of the transition from Horizontal Layering to Vertical Integration as an organizational response to AI capabilities. Three key findings emerge:

First, we observe structural consolidation, wherein traditional functional silos (frontend, backend, testing, operations) collapse into integrated Super-Cells. A single AI-augmented engineer can now maintain end-to-end ownership of feature lifecycles that previously required cross-functional teams.

Second, we document substantial efficiency gains. Project W demonstrated an 8.3$\times$ reduction in resource consumption (12 vs. 100 person-months), while the Internal AI-CRM project achieved a 33$\times$ reduction (1.5 vs. 50 person-months). These order-of-magnitude improvements validate the theoretical premise that AI fundamentally alters the economics of software production.

Third, we identify cognitive role transformation. Engineers transition from execution-focused coding to high-level architectural supervision and liability governance, consistent with our Cognitive Bandwidth Optimization theory.

\subsection{Theoretical Contributions}

This study advances three theoretical propositions:

We introduce Human-AI Collaboration Efficacy as the normative standard for evaluating engineering organizations in the AI era, replacing traditional metrics focused on individual productivity or output volume.

We identify the AI Distortion Effect within Total Factor Productivity dynamics, demonstrating that AI amplifies technological contributions while suppressing the marginal utility of labor scale. This finding has significant implications for workforce planning and organizational sizing.

We extend Conway's Law to the AI context, showing that reduced organizational nodes (N) correlate with reduced system complexity, but that this simplification transfers cognitive and liability burdens to individual human supervisors.

\subsection{Managerial Implications}

For practitioners, our findings suggest three strategic imperatives:

Organizations should restructure around Type II Super Employees, engineers capable of leveraging AI for exponential efficiency gains within their domains, rather than maintaining rigid functional departments.

Traditional artifacts such as Product Requirement Documents and formal handover protocols constitute digital waste in the AI paradigm and should be minimized or eliminated.

Human-in-the-Loop governance mechanisms are essential as efficiency increases, ensuring that human judgment remains the final arbiter of system integrity and liability.

\subsection{Limitations}

Several limitations qualify our findings. First, elite bias and cognitive sustainability present concerns: our study relies on high-performing engineers who successfully transitioned to Type II Super Employees, and it remains unclear whether median-level developers can replicate these outcomes. The sustained cognitive intensity of the Super-Cell structure may also induce burnout over periods longer than our 3-month observation window.

Second, methodological confounds affect interpretation. The Seed Team's awareness of their pilot status likely contributed to the Hawthorne Effect, and scaling from a motivated 4-person unit to enterprise-scale organizations involves cultural resistance not captured here. Additionally, Organization B's dual role as research partner and case subject introduces potential observer bias despite triangulation with objective metrics.

Third, long-term technical debt remains unexamined. While Vibe Coding \citep{ray2025} accelerated delivery, the maintainability of AI-generated codebases over 3-5 year lifecycles is untested. Should original prompts be lost or underlying LLM capabilities drift, future maintainers may face illegible black-box logic.

Finally, model dependency constrains generalizability. The observed efficiency gains (8$\times$--33$\times$) are contingent on current state-of-the-art LLM capabilities; degradation in model performance or shifts in API availability could render the Vertical Integration workflow economically unviable.

\subsection{Future Research Directions}

Future research should examine longitudinal maintainability of AI-generated systems over 3-5 year lifecycles, scalability of the Super-Cell model to ultra-large distributed systems, cognitive sustainability and burnout risks under sustained high-load supervision, and independent replication across diverse industry contexts.

In conclusion, the software industry is transitioning from a labor-intensive to an intelligence-intensive discipline. Organizations that successfully navigate this transition will be those that reconstruct their structures to optimize human-AI collaboration rather than merely deploying AI as an incremental productivity tool.

\section*{Declarations}

\textbf{Author Contributions:} Zhang C., Li Z., and Zhong Z. conceptualized the study, provided the primary empirical data, and formulated the core theoretical framework based on their practice at Organization B. Ma H. designed the qualitative research methodology and conducted the in-depth interviews. Xiao D. and Lin C. were responsible for figure design and copy-editing. Dong M. provided critical academic advice and validated the research logic. All authors reviewed and approved the final manuscript.

\textbf{Competing Interests:} Zhang C., Li Z., Zhong Z., Xiao D., and Lin C. are full-time employees of Organization B, which served as the case study subject and research collaborator. Ma H. serves as the Chief Scientist (part-time) of Organization B, in addition to his primary academic affiliation. While the research is grounded in the company's practices, the authors adhered to academic standards of objectivity in data analysis. Dong M. served as an independent academic consultant and declares no employment ties to Organization B. Data collection and analysis were conducted following standard qualitative research protocols to minimize potential bias arising from organizational affiliation.

\textbf{Generative AI Declaration:} Generative AI tools were used for copy-editing, language refinement, and formatting assistance during the preparation of this manuscript. The authors have reviewed all AI-generated output and take full responsibility for the content, accuracy, and originality of the work.

\bibliographystyle{apalike}
\bibliography{references}

\end{document}